\documentclass[fleqn,usenatbib,useAMS]{mnras}

\usepackage{newtxtext,newtxmath}
\usepackage[T1]{fontenc}
\usepackage{ae,aecompl}

\usepackage{graphicx}	% Including figure files

\usepackage{units}
\newcommand{\eq}[1]{\begin{equation} #1 \end{equation}}

\title[Can EDGES favour any dark matter model?]{Can EDGES observation favour any dark matter model?}

\author[A. Rudakovskyi et al.]{
A. Rudakovskyi,$^{1,2}$\thanks{E-mail: rudakovskyi@bitp.kiev.ua}
D. Savchenko,$^{1,2}$
M. Tsizh$^{3}$
\\
$^{1}$Bogolyubov Institute for Theoretical Physics of the NAS of Ukraine, Metrolohichna Str. 14-b, Kyiv, 03143, Ukraine\\
$^{2}$Kyiv Academic University, 36 Vernadsky blvd., Kyiv, 03142, Ukraine\\
$^{3}$Astronomical Observatory of Ivan Franko National University of Lviv, Kyryla i Methodia Str., 8, Lviv 79005, Ukraine
}

\date{Accepted XXX. Received YYY; in original form ZZZ}

\pubyear{2020}

\begin{document}
\label{firstpage}
\pagerange{\pageref{firstpage}--\pageref{lastpage}}
\maketitle

\begin{abstract}
The recent detection of the 21-cm absorption signal by the EDGES collaboration has been widely used to constrain the basic properties of dark matter particles. However, extracting the parameters of the 21-cm absorption signal relies on a chosen parametrization of the foreground radio emission. Recently, the new parametrizations of the foreground and systematics have been proposed, showing significant deviations of the 21-cm signal parameters from those assumed by the original EDGES paper. In this paper, we consider this new uncertainty, comparing the observed signal with the predictions of several dark matter models, including the widely used cold dark matter model, 1--3~keV warm dark matter models, and 7~keV sterile neutrino (SN7) model, capable of producing the reported 3.5~keV line. We show that all these dark matter models cannot be statistically distinguished using the available EDGES data.
\end{abstract}


%%%%%%%%%%%%%%%%%%%%%%%%%%%%%%%%%%%%%%%%%%%%%%%%%%

%%%%%%%%%%%%%%%%% BODY OF PAPER %%%%%%%%%%%%%%%%%%

\section{Introduction}
The possibility of observation of the cosmological 21-cm hydrogen signal was of interest to cosmologists even before the first observational constraints became real \citep[see e.g.][]{Hogan:1979, Scott:1990,Madau:1996cs}. 
There are two possible ways in which this observation may shed light on the structure formation processes in the early Universe. The first one is the 21-cm tomography, which is based on the study of the spatial distribution of fluctuations of the 21-cm signal generated by HI clouds at the Dark Ages and reionization epochs \citep[see e.g.][]{Madau:1996cs,Ciardi:2004ru,McQuinn:2005hk,Mao:2008ug,Morales:09}. 
The second one is the detection of the sky-averaged signal from the Dark Ages epoch~\citep[see e.g.][]{Mirocha:13,Mirocha:2015jra,Cohen:2016jbh} produced by absorption of the cosmic microwave background (CMB) radiation by the neutral hydrogen. This absorption is caused by the Wouthuysen--Field coupling between the spin temperature of hydrogen and Ly~$\alpha$ radiation of the first galaxies. For a detailed description of the theoretical and observational challenges of 21-cm signal detection,~\citep[see e.g.][and references therein]{Furlanetto:2006jb, Pritchard:2008da, Pritchard:2011xb}.

Detection of the global absorption signal claimed by the EDGES collaboration~\citep{Bowman:2018yin} caused great excitement among physicists. While the central frequency of the absorption peak (which is linked to the redshift at which it was generated) is in a good agreement with the predictions of the Lambda--CDM model~\citep{Cohen:2016jbh},\footnote{However, explanation of the frequency of absorption feature reported by EDGES requires more efficient star formation in low-mass galaxies compared to that extrapolated from $z \sim 6-8$; see more in \citep{Mirocha:2018}.} the amplitude ($\sim0.5$\,K) of the observed signal appears to be at least twice that of the most extreme absorption predicted in the Lambda--CDM model \citep[see e.g.][]{Cohen:2016jbh}.

Observation of the global 21-cm absorption by EDGES motivated a wide search of possible mechanisms that would explain the observed depth and position of the absorption signal. This opened up a possibility to study and put constraints on the early star formation rate \citep[see e.g.][]{Madau:18, Schauer:19}, structure formation in dark matter models \citep{Safarzadeh:2018hhg,Schneider:2018xba,Chatterjee:19,Boyarsky:2019fgp,Leo:19} and non-standard X-ray sources that can heat the IGM such as the first black holes \citep{Clark:2018ghm} and decaying or annihilating dark matter \citep{Mitridate:2018iag, Cheung:2018vww, Liu:2018uzy, Yang:2018gjd,DAmico:2018sxd,Fraser:2018acy, Hektor:18,Clark:2018ghm, Chatterjee:19, Chianese:19}. Additional mechanisms beyond the standard cosmological scenario were proposed to explain the depth of the absorption signal, such as the interaction between baryonic matter and dark matter \citep{Barkana:2018lgd, Fialkov:2018xre, Bhatt:19} and extreme radio background during the Dark Ages \citep{Feng:18,Ewall-Wice:18, Fialkov:19a}, alternative dark energy models \citep{Hill:18, Li:19,Yang:19}.  

At the same time, a well-known problem of detecting the 21-cm signal is that the galactic synchrotron emission and ionospheric emission and absorption are dominant over the signal by four orders of magnitude at frequencies below 100 MHz~\citep{Bernardi:2014caa}. Therefore, it is not surprising that soon after the EDGES publication several works appeared addressing the technical details of extracting the 21-cm signal from the observation and questioning the reliability and validity of the observed profile. \citet{Hills:2018vyr} reanalysed the data reported by EDGES. They showed that the original analysis assumed a controversial structure in the spectrum of the foreground (galactic) emission in the studied spectral band and led to non-physical properties of the ionosphere. \citet{Bradley:2018eev} reported a  possible systematic artefact in the observations that can
affect the determination of the absorption signal. 

The 21-cm signal naturally depends on  structure formation history. In turn this history depends on the underlying dark matter model. Unlike the standard CDM scenario, in warm dark matter (WDM) models particles have masses of the order of few keV and relativistic initial velocities. This fact leads to the smearing of density perturbations on scales below the free-streaming length $\lambda_\text{fs}$~\citep{Boyarsky:08,Boyarsky:2018a} and suppression of the structure formation on such small scales, compared to the predictions of CDM. The WDM particle candidates naturally arise in some extensions of the Standard Model [gravitino~\citep[see e.g.][]{Viel:05}, sterile neutrino~\citep[see e.g.][and references therein]{Boyarsky:2018a} etc.]. The warm dark matter particles may reach thermal equilibrium with other particles (gravitinos) or have non-equilibrium distribution (sterile neutrinos).
The lack of the small-scale structures in the warm dark matter scenario could modify the position and the form of the 21-cm signal in comparison with CDM \citep[see e.g.][]{Sitwell:13}. The main goal of this manuscript is a study of the possibility to distinguish between warm and cold DM models via the `raw' EDGES data.

This work has the following plan. In the next section we describe how we use the foreground modelling proposed in \citep{Hills:2018vyr} and take into account the ground plane artefact discussed in \citep{Bradley:2018eev} to best fit the EDGES data via the non-linear least-squares procedure. Next, we model the 21-cm global signal with the \texttt{ARES} code for different models of dark matter [CDM, thermal relic WDM and 7 keV sterile neutrino decaying dark matter motivated by the recently detected 3.5\,keV line \citep{Boyarsky:2014jta,Bulbul:2014sua, Boyarsky:2018ktr}\footnote{Such dark matter is in a good agreement with the Ly~$\alpha$ forest analysis \citep{Baur:2017stq} and reionization history data \citep{Rudakovskyi:2018jfc}.}]. Then, in the \textit{Results} section we fit the EDGES data subtracting the obtained different absorption signals and compare the fitting scores in order to select the preferred dark matter model. In the last, \textit{Conclusions and Discussion} section we briefly summarize our results and discuss the possibility to constrain the parameters of structure formation during the Dark Ages and, in particular, dark matter properties by the EDGES data and by the planned observations of 21-cm absorption signal.

\section{Methods}\label{sec:methods}
\subsection{Fitting the EDGES data}\label{sec:methods-fit}
We represent the EDGES data with a sum of three components, 
\begin{equation}
T(\nu) = T_\text{sky}(\nu) + T_\text{res}(\nu) + T_\text{21}(\nu). \label{eq:T-nu}
\end{equation}

Here, the first term is the sky foreground brightness temperature. In the original EDGES paper~\citep{Bowman:2018yin}, this foreground was modelled in the linearized form. However,~\citep{Hills:2018vyr} argued that the parameters of such a representation assume non-physical values in the best-fitting model. 
In our fits, we used the non-linear model with the ionospheric absorption and emission terms being connected through the electronic temperature $T_\text{e}$, obtained by expanding the foreground temperature around some central frequency $\nu_\mathrm{c}$ as described in~\citep{Hills:2018vyr}:
    \begin{multline}
    T_{\rm sky}^{\text{H18}}(\nu) =  b_0\left
(\frac{\nu}{\nu_c}\right)^{-2.5+b_1+b_2\ln(\nu/\nu_c)}\mathrm{e}^{-b_3(\nu/\nu_c)^{-2}} +\\
+ T_{\rm e}\left(1-\mathrm{e}^{-b_3(\nu/\nu_c)^{-2}}\right).
\label{eq:Tsky}
    \end{multline}
Here, the electronic temperature $T_e$ is allowed to vary in the range [200, 2000]~K, according to the EDGES measurements, see fig.~2 of~\citep{Rogers:14}. 
The parameter $b_1$ is a correction to the overall power-law index, which varies by $\sim$0.1 across the sky according to~\citep{Bowman:2018yin}. In our fits we allow this parameter to vary in a broader range of $[-0.2, 0.2]$. The parameter $b_2$ was left unconstrained, although its best-fitting value should be controlled to be $\sim$0.1 according to~\citep{Bernardi:2014caa}. The ionospheric opacity $b_3$ is allowed to vary between 0 and 0.03 according to~\citep{Hills:2018vyr}.
    
The second term in equation~(\ref{eq:T-nu}) is the sum of three resonant absorption components described in ~\citep{Bradley:2018eev}:
\eq{T_{\rm res}(\nu) = -\sum\limits_{i=1}^{3} \frac{A_{i} \nu^3 \nu_{0i}}{\nu^4 + Q_i^2  (\nu^2 - \nu_{0i}^2)^2}.\label{eq:Tres}}
This term is characterized by three parameters: the central frequencies $\nu_{0i}$,  the depths of the profiles $A_i\equiv A(\nu_{0i})$, and the quality factors $Q_i$  which are the ratios of $\nu_{0i}$ to the spectral widths of  absorption.
The phenomenological origin of this term is the fact that soil itself is a resonator, of which one cannot get rid. EDGES uses the ground antennae, and possible discontinuities at the edges of the ground plate produce resonant features and distort the signal.

The last term in equation~(\ref{eq:T-nu}), $T_\text{21}(\nu)$, is the global 21-cm absorption signal. While~\citep{Bowman:2018yin} modelled it in the form of flattened Gaussian, we do not represent it in any analytical form. Instead, from the initial EDGES data we subtract the absorption profile obtained in the \texttt{ARES} simulation, taking into account the specific structure formation model, which includes the influence from the particular type of the halo mass function and DM decays (see Sec.~\ref{sec:ares}).

The EDGES data do not include error bars. Therefore, we use the two following approaches for analysis of the EDGES data.
According to the first one, we follow~\citet{Bradley:2018eev} and choose the objective function in the form of the minimal log likelihood $ -\mathrm{ln}\,L = \frac{1}{2} \sum_{kl} (y_k - \hat y_k)C^{-1}_{kl}(y_l - \hat y_l)$
with the noise covariance matrix $C_{kl}\propto y_k^2\delta_{kl}$, where $y_i$ are the data points and $\hat y_i$ are the model values. The constant of proportionality is not defined; however, it changes only the magnitude of the log-likelihood as function of parameters, not the shape, allowing to produce a weighted fit.  
As a result of the fitting procedure, the root mean square (rms) score could be calculated for the best-fitting parameters. Due to the undefined absolute values of the errors, we cannot use the standard chi-squared test to quantify the differences in the goodness-of-fits between models.

Therefore, in the second approach we follow the logic of~\citet{Singh:19} and assume the error bars to be the same for all data points. The covariance matrix has the form $C_{kl}=c_k\delta_{kl}$ here, with the constant of proportionality $c_k$ being also unknown. This leads to the unweighted least-squares fitting procedure.
In such case, we can use the Bayesian information criterion (BIC) for model selection. For the uniform Gaussian errors it could be written as~\citep{Schwarz:78,Priestley:83}
\begin{equation}
    \text{BIC} = N\ln \left(\frac1N \sum_{i=1}^{N} (y_i - \hat y_i)^2\right) +k\ln N\,.
\end{equation}
Here, $y_i$ denotes the data points and $\hat y_i$ are the model values, $N$ is the number of the data points ($N=123$ for the EDGES data), and $k$ is the number of free model parameters ($k=15$ for our model).
Notice that the extraction of the simulated signal from the raw spectra does not change the values of $N$ and $k$.
The model with the lower BIC could be treated as strongly preferred if the difference between BICs is higher than 6, and one cannot talk of any preference for $\Delta\text{BIC}<2$~\citep{Robert:95}.

In both case, we use the non-linear least-squares procedure implemented in the \texttt{lmfit PYTHON}  package~\citep{lmfit}. 
Because the underlying Levenberg--Marquardt algorithm depends on the initial guess for the parameters, we repeat fits many times, randomly varying these initial values and select the fit with the highest likelihood.
\subsection{Modelling the global 21-cm absorption profile}\label{sec:ares}
We use the open-sourced Accelerated Reionization Era Simulations (\texttt{ARES}) code~\citep{Mirocha:2014faa, Mirocha:16, Mirocha:2018} to compute the global 21-cm neutral hydrogen signal in different DM models. This code produces the profile of the global 21-cm signal for given star formation rate, halo mass function and cosmology. The PopII stars in the galaxies are assumed to be the sources of the Ly~$\alpha$ and ionizing UV-radiation, and black holes produce X-rays. Also, there is a possibility to add new sources of radiation, like, for example, decaying DM.

We adopt the cosmological parameters defined by~\citet{Planck:2015xua}, namely $\Omega_{\Lambda}=0.685$, $\Omega_m=0.315$, $\Omega_b=0.049$, $h=0.673$, $n_s=0.965$, and $\sigma_8=0.83$.

The halo abundance is usually described by the halo mass function:
\eq{\frac{\mathrm{d}n}{\mathrm{d}\mathrm{ln }M}=f(x)  \frac{\bar{\rho}_{m}}{M}\frac{\mathrm{d}\mathrm{ln}\sigma^{-1}}{\mathrm{d}\mathrm{ ln} M},}
where  $x=\left(\frac{\delta_c^2(z)}{\sigma^2}\right)$, $\delta_c(z)=\frac{1.686}{D(z)}$, $D(z)$ is the growth factor \citep{Heath:1977}, $\sigma(M)$ is the variance of the density fluctuations on the mass scale $M$, and $\bar{\rho}_{m}$ is the mean matter density of the Universe.

We use the results of the $N$-body simulations for thermal relic WDM and CDM scenarios at high redshifts from~\citet{Schneider:2018xba} as a reference for our halo mass function fitting formula. 
Following~\citet{Schneider:2018xba} for all DM models, we use the Sheth--Tormen approximation \citep{Sheth:01}: \eq{f(x)=A_\textit{ST}\sqrt{\frac{2qx}{\upi}}\left(1+(qx)^{-p}\right) \mathrm{e}^{-qx/2}.} 
We find that the parameters $A_\textit{ST}=0.322$, $q=0.93$, and $p=0.3$ are in a best agreement with the CDM and WDM simulations, shown in fig.1 of \citep{Schneider:2018xba}\footnote{Note that in \citet{Schneider:2018xba} the parameter $q$ is claimed to be equal to~1. However, we found that $q=0.93$ is the best-fitting parameter value for the simulation points provided in \citet{Schneider:2018xba}.
}.

Following \citet{Schneider:2018xba}, we also consider a sharp-$k$ filter for $\sigma(M)$ calculation, which provides a good fit for the CDM and WDM halo mass functions, obtained during $N$-body simulations at different redshifts \citep{Benson:12,Schneider:14, Schneider:2018xba}: $\sigma^2=\int^{k_c}_{0}P(k)\frac{\mathrm{d}^3k}{(2\upi)^3}$, where $P(k)$ is the dark matter power spectrum, which depends on the properties of the DM particles.  
The mass of the halo and $k_c$ are related as $M=\frac{4\upi}{3}\bar{\rho}_m\left(\frac{2.5}{k_c}\right)^3$\!. \

Warm dark matter power spectrum can be connected with cold dark matter power spectrum $P_\text{CDM}(k)$ as: $P_{\text{WDM}}(k)=P_{\text{CDM}}T^2(k)$, where $T(k)$ is the so-called transfer function. For transfer functions of WDM in the form of thermal relic (e.g. gravitino), we adopt the fitting formula by~\citet{Viel:05}:
\begin{equation}
    T(k)=(1+(\alpha k)^{2\nu})^{-5/\nu},
\end{equation}
where
\eq{\alpha=0.049\left(\frac{m}{1\text{~kev}}\right)^{-1.11}\left(\frac{\Omega_{\text{WDM}}}{0.25}\right)^{0.11}\left(\frac{h}{0.7}\right)^{1.22}h^{-1}\text{~Mpc},}
$\Omega_{\text{WDM}}=\Omega_\text{DM}$, $\nu=1.12$, $m$ is the mass of thermal relic WDM particle.

While in WDM models in the form of thermal relic the free-streaming length and power spectrum  depend only on the particle mass, for sterile neutrino they also depend on the production mechanism. We use power spectrum for the 7~kev sterile neutrino, generated with the lepton asymmetry $L_6=12$ \citep{Lovell:2015psz,Lovell:2016fec}.

\begin{figure}
    \includegraphics[width=\columnwidth]{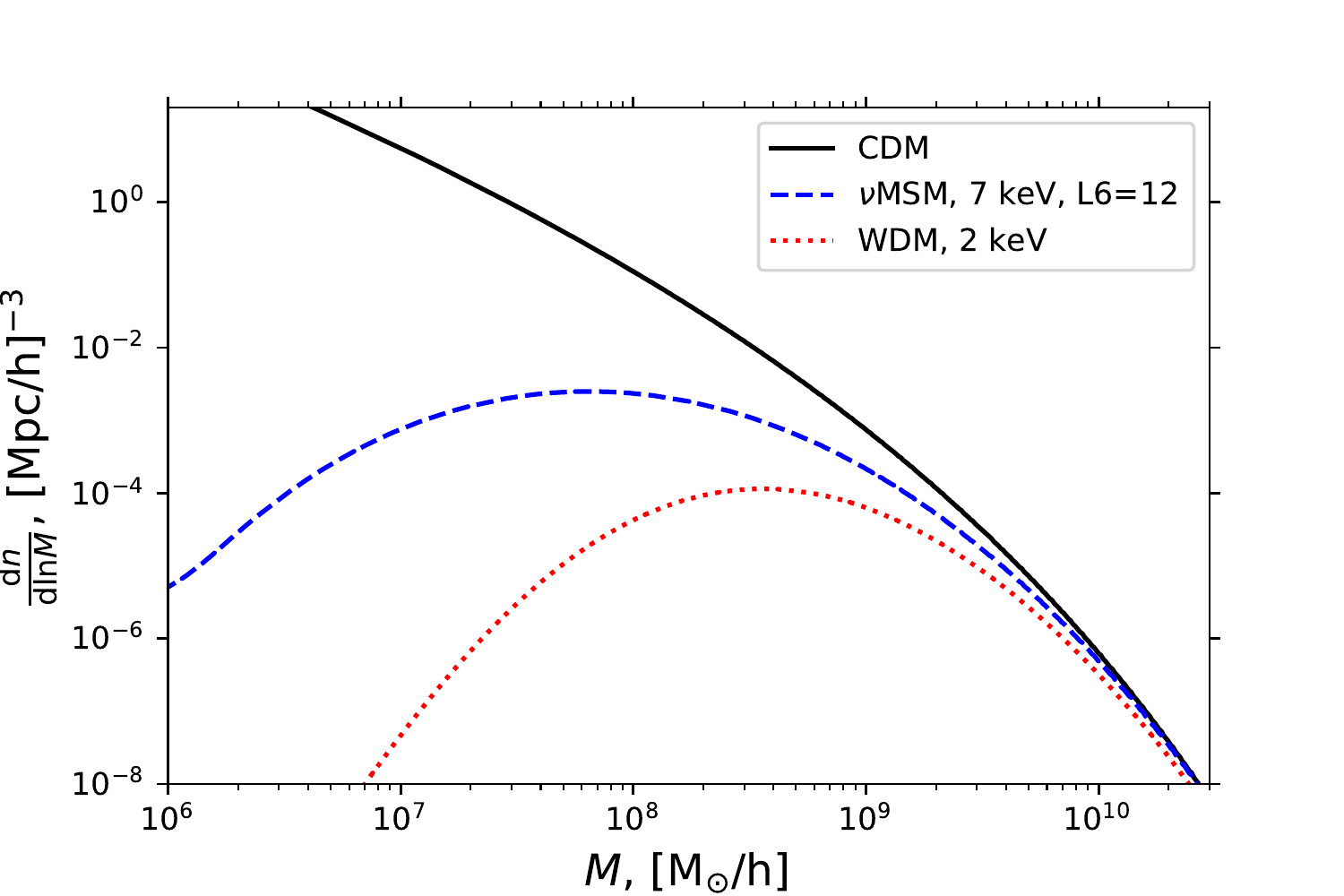}
    \caption{Halo mass functions for CDM, WDM with particle mass $m=2$\,keV, and 7\,keV sterile neutrino with $L_6=12$ at $z=17$.}
    \label{fig:hmf}
\end{figure}

For each of the DM models under consideration, we generate the corresponding halo mass functions for redshifts $z$ between 0 and~50. For thermal relic WDM and CDM we use the \texttt{hmf} public code \citep{Murray:2013qza}. Examples of such halo mass functions at $z=17$ are shown in Fig.~\ref{fig:hmf}.

The star formation rate density (SFRD) is modelled by the standard so-called fcoll model used in \texttt{ARES}: \eq{\dot\rho_*=f_*\rho_\text{b}\frac{\mathrm d}{\mathrm d t}f_\text{coll}(z).}

The star formation efficiency $f_*$ in our analysis is considered, for simplicity, to be a constant not exceeding unity. The constant star formation efficiency is different from the default \texttt{ARES} parametrization of $f_*$ by a double-power-law function of the halo mass \citep{Mirocha:16}. The plausibility of our assumption is motivated by the fact that the star formation efficiency is unknown at high redshifts, and by a degeneracy between the effects of mass of the particle of warm dark matter and $f_*$ \citep[see e.g.]{Sitwell:13, Boyarsky:2019fgp}. The collapsed fraction $f_\text{coll}$ is defined as
\eq{f_\text{coll}(z)=\frac{1}{\rho_m}\int^{\infty}_{M_\text{min}}\frac{\mathrm{d}n}{\mathrm{d}\mathrm{ln} M}\mathrm{d}M,}
where $M_\text{min}$ is the minimal mass of the source, determined by the virial temperature $T_\text{vir}$, similarly to \citep{Barkana:2000fd}. In this work we assume that $T_\text{vir}=10^4K$.
The specific emissivity in a particular spectral band between $E_\mathrm{min}$ and $E_\mathrm{max}$ is proportional to \textit{SFRD} \cite[see e.g.][]{Mirocha:2014faa}:
\eq{\epsilon(E,z)=c \dot\rho_*(z) I(E),}
where $c$ is the conversion factor between emissivity and \textit{SFRD}, and $I(E)$ is the spectral density normalized as  $\int^{E_\mathrm{max}}_{E_\mathrm{min}}I(E)\mathrm{d}E=1$.

Throughout this work, we assume the default \texttt{ARES} parameters (except for $f_\star$)  for the Ly~$\alpha$ and Ly~C photons produced by PopII stars in the first galaxies and X-rays generated by the accretion of baryonic matter onto the first black holes.

Another possible source of X-ray photons are decaying DM particles. The recently detected narrow 3.5~keV line in the spectra of the DM dominated objects \citep{Boyarsky:2014jta,Boyarsky:2014ska,Bulbul:2014sua} may be emitted during the DM decay. One of the best-motivated DM candidates, which might explain the 3.5~keV line, is $\sim 7$~keV sterile neutrino~\citep[see e.g.][and references therein]{Adhikari:2016bei,Boyarsky:2018a}. We focus on the resonantly produced $7$~keV sterile neutrino DM with the lepton asymmetry $L_6=12$, corresponding to the mixing angle $\sin^22\theta= 1.6\times 10^{-11}$ according to \citep{Lovell:2015psz}, and lifetime $\tau_\text{DM}=2.677\times 10^{28}$~s consistent with \citet{Boyarsky:2018ktr}.
We model the dark matter decays in \texttt{ARES} as a new population similar to black holes with constant accretion rate. For decaying DM models (including the model of 7~keV sterile neutrino), the luminosity density of photons with energy $E_\gamma$ generated via decays is
\eq{L_\text{decays}=E_{\gamma}\frac{\rho_\mathrm{DM}}{m_\mathrm{DM}}\frac{1}{\tau_\mathrm{DM}}.}

\vspace{11pt}
\section{Results}

First, we basically reproduce the modelling by \citep{Bowman:2018yin} and \citep{Bradley:2018eev} to justify our fitting method (see Appendices~\ref{appendix:bowman} and~\ref{appendix:bradley}).
 
Then we generate the absorption signals for the different dark matter models (CDM; 1, 2, and 3~keV thermal relic WDM, 7~keV decaying sterile neutrinos), each with different values of $f_*$ in the range from 0.01 to 1, while all other astrophysical and cosmological parameters are fixed. We subtract these signals from the data and perform the fits as described in Sec.~\ref{sec:methods-fit}. 

The minimal rms in the weighted least-square procedure is obtained for the 3~keV WDM model with $f_\star=0.52$ and is equal to $20.72$~mK and the maximal rms is $20.9$~mK in the 2 keV WDM model with $f_\star=0.8$. All these values are close to $20.8$~mK obtained in \citep{Bradley:2018eev} and better than $25$~mK from \citep{Bowman:2018yin}. However, one cannot draw robust conclusions about the difference in the fitting qualities between these models due to the different number of free parameters in models.

In the unweighted least-square approach, the obtained rms for all DM models and all $f_\star$ are in the range of $20.66-20.72$~mK, which correspond to BICs in the range of $-881.47$ to $-882.20$ for $k=15$ and $N=123$. Thereby, differences in the values of BICs are all below 2, so we conclude that all tested dark matter particle models are indistinguishable.

\begin{figure*}
    \centering
    \includegraphics[width=0.85\textwidth]{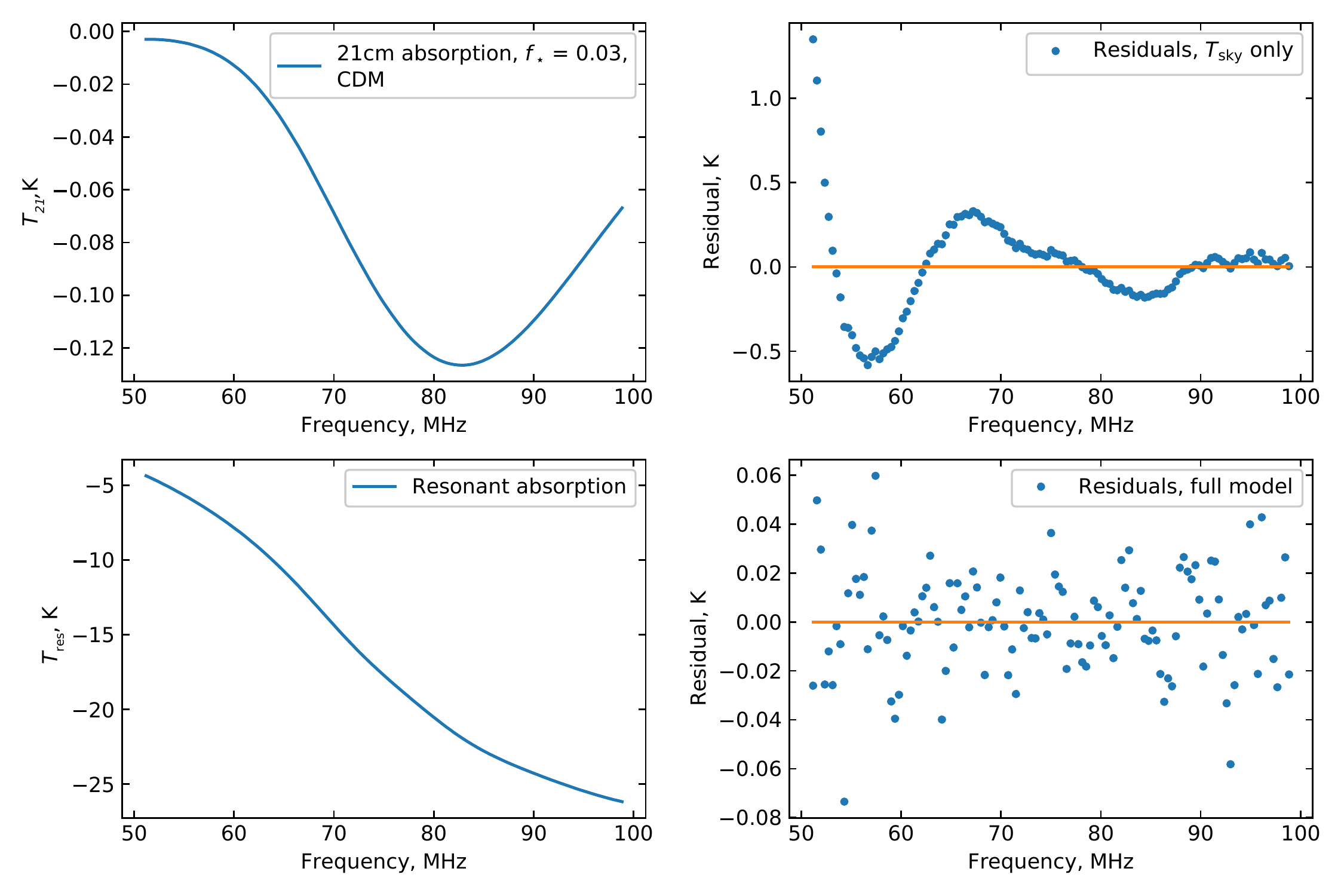}
    \caption{Results of the weighted fits,  assuming the 21-cm absorption in the CDM model with $f_\star=0.03$.  \textit{Top left:}~Absorption profile in the frequency range where the fit is performed. 
    \textit{Top right:}~Residuals after fitting and removing only the foreground model, equation~(\protect\ref{eq:Tsky}).
    \textit{Bottom left:}~The best-fitting model of the instrumental resonant absorption feature, equation~(\protect\ref{eq:Tres}).
    \textit{Bottom right:}~Residuals after removing both the foreground and the resonant absorption model.}
    \label{fig:cdm}
\end{figure*}

\begin{figure*}
    \centering
    \includegraphics[width=0.85\textwidth]{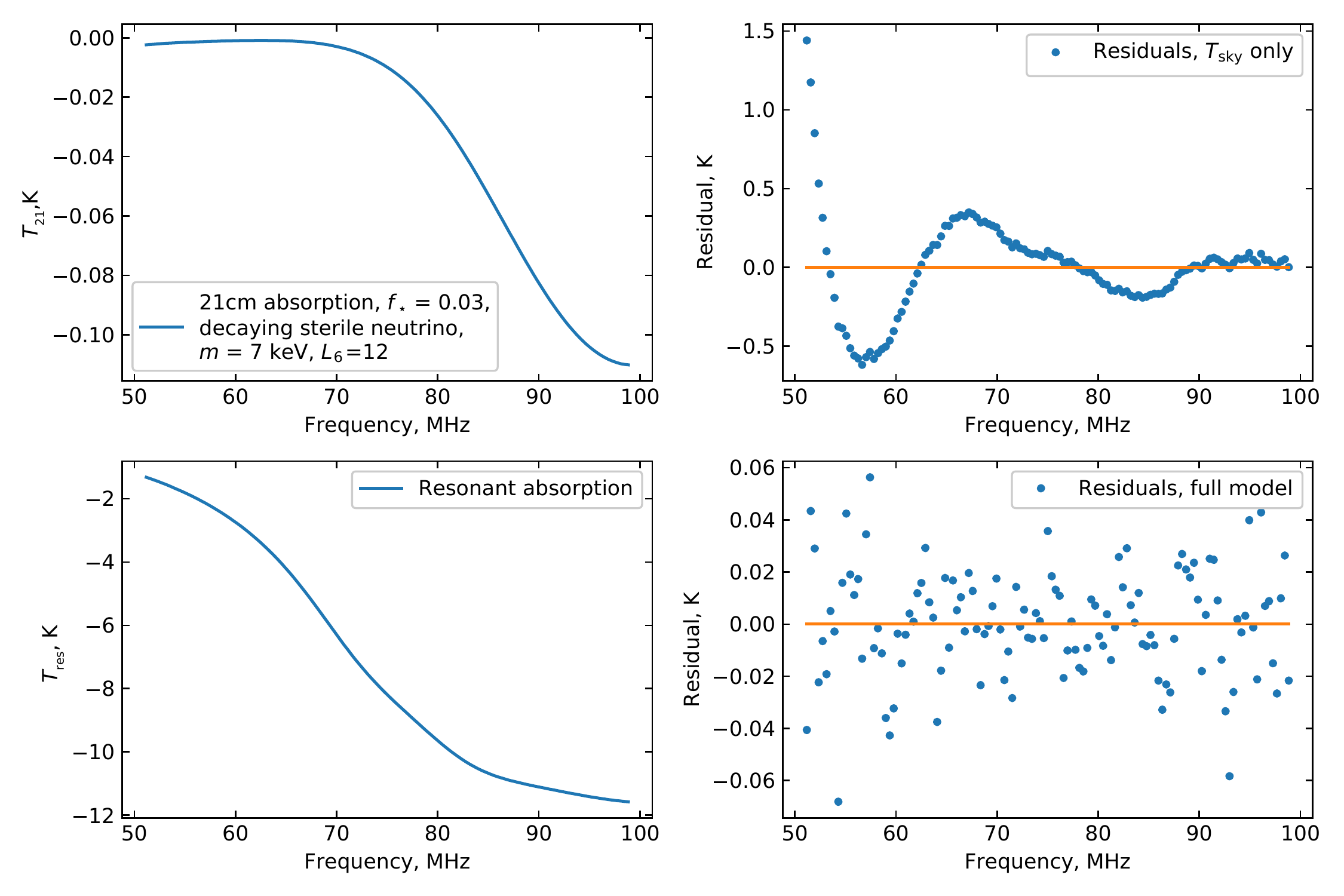}
    \caption{The same as in Fig.~\protect\ref{fig:cdm}, but assuming the 21-cm absorption in the 7\,keV sterile neutrino dark matter model with $f_\star=0.03$.}
    \label{fig:7kev-dec}
\end{figure*}

Examples of specially interesting cases (standard CDM and decaying sterile neutrino dark matter) are shown in Figs.~\ref{fig:cdm} and~\ref{fig:7kev-dec}. 

We also test the case without any absorption signal at all. In the weighted least-squares procedure the obtained rms is $20.87$~mK. In the unweighted approach the rms is $20.7$~mK, which corresponds to $\text{BIC}=-881.7$, showing that the absence of a signal is also allowed by the data.

Additionally, we perform unweighted least-square fits with the foreground model alone, not including the instrumental resonant components (number of parameters $k=6$). Subtracting signals, corresponding to the CDM and 2~keV WDM, we find, that the best-fitting rms is $0.215$~mK, corresponding to $\text{BIC}=-349.09$. Moreover, residuals have a clearly visible oscillating feature. This shows that foreground-only model is strongly disfavoured.

We have performed the same computations with Planck-18 cosmological parameters \citep{Planck:18}. In such case, the calculated rms and BICs appear to be very close to the results, obtained in Planck-15 cosmology.

\section{Conclusions and Discussion}

We have modelled the EDGES data with the alternative to \citep{Bowman:2018yin} physically motivated foreground model  proposed by~\citet{Hills:2018vyr}, taking into account the ground plane artefact absorption by~\citet{Bradley:2018eev}, and subtracting different simulated 21-cm absorption profiles in order to constrain the underlying dark matter model. We have explicitly shown that the fit quality of such model does not depend on the assumed dark matter particle model, concluding that the EDGES observation cannot be used as a good tool for the quantitative constraining of dark matter particle models. Moreover, one cannot even distinguish between existence or absence of a signal.

Unlike the papers in which the form and position of the signal reported by~\citep{Bowman:2018yin} are used to constrain the dark matter models \citep[see e.g.][]{Safarzadeh:2018hhg, Schneider:2018xba,Clark:2018ghm, Hektor:18, Chatterjee:19}, our paper uses the `raw' signal $T(\nu)$ to perform the modelling. 

\citep{Bradley:2018eev} proposed a physically motivated instrumental feature; however, one can try to check other forms of systematics. For example, \citep{Singh:19} showed that the EDGES spectrum is consistent with the standard cosmology if the maximally smooth polynomial foreground and sinusoidal systematics are assumed. Their fit gives  $\text{BIC}=-894.2$, formally strongly preferred over our best value of $-882.2$. It is a matter of choosing the form of the foreground and of the term describing the systematics. This brings in uncertainty and model dependence into the exploration of the global absorption signal. Moreover, the global 21-cm signal is averaged over the sky and contains contributions from many different sources; thus no simple physical model may be appropriate to fit it.

\citet{Sims:19} also make direct fitting of EDGES data considering \texttt{ARES} simulations of the 21-cm signal and use the oscillating form of systematics. Direct parametrization of the noise covariance matrix allows them to explicitly evaluate the log(evidence) in the weighted fitting procedure.
However, only standard cold dark matter scenario have been considered.

The developed radiometer experiments such as BIGHORNS \citep{Sokolowski:15}, SARAS 2 \citep{Singh:17}, and LEDA \citep{Price:17}, which will be focused on the global 21-cm  signal,  may shed new light on the 21-cm absorption feature.
However, there are large uncertainties in the star formation in galaxies during the reionization and Dark Ages epochs, which makes it difficult to constrain a dark matter scenario by using the global 21-cm absorption signal \citep{Boyarsky:2019fgp}.
Nevertheless, the future studies of the statistics of the spatial distribution of the 21-cm signal by radio interferometers such as MWA, HERA
and SKA may help to break the degeneracy between the baryonic and dark matter effects during the reionization and Dark Ages epochs \citep[see e.g.][]{Mesinger:2013, Sitwell:13, Bull:2018lat}.

\section*{Acknowledgements}

The authors are grateful to Dr. D.~Iakubovskyi and Dr. G.~Sun for valuable comments and to Prof. Dr. Yu.~Shtanov for reading and commenting on this paper. We  thank  the anonymous referee for the comments that significantly improved the quality of the paper. This work was supported by the grant for young
scientist research laboratories of the National Academy of Sciences of Ukraine. The work of A.R.
was also partially supported by the ICTP through AF-06.

\appendix
\section{Reproducing the original EDGES results}\label{appendix:bowman}
In the original EDGES paper \citep{Bowman:2018yin}, the foreground was modelled as
\begin{multline}\label{eq:bowman-foreground}
T_\text{sky}^{\text{(B18)}}(\nu)=a_0 \left(\frac{\nu}{\nu_c}\right)^{-2.5}+a_1 \left(\frac{\nu}{\nu_c}\right)^{-2.5} \ln\left(\frac{\nu}{\nu_c}\right) +\\
+ a_2 \left(\frac{\nu}{\nu_c}\right)^{-2.5} \left[\ln\left(\frac{\nu}{\nu_c}\right)\right]^2 + a_3 \left(\frac{\nu}{\nu_c}\right)^{-4.5} + a_4 \left(\frac{\nu}{\nu_c}\right)^{-2}.
\end{multline}
The absorption signal was assumed in the form of a flattened Gaussian:
\begin{equation}\label{eq:bowman-flgauss}
    T_{21}^{(\text{B18})}(\nu) = -A\left( \frac{1-\mathrm{e}^{-\tau\mathrm{e}^B}}{1-\mathrm{e}^{-\tau}} \right),
\end{equation} where
\begin{equation}
    B=\frac{4(\nu-\nu_0)^2}{w^2}\ln\left[ -\frac{1}{\tau}\ln\left(\frac{1+\mathrm{e}^{-\tau}}{2}\right)\right],
\end{equation}
 We fit the data with the sum of the sky foreground brightness temperature in the form of equation~(\ref{eq:bowman-foreground}),
and the 21-cm absorption term in the form of the flattened gaussian, equation~(\ref{eq:bowman-flgauss}). 
\begin{figure}
    \centering
    \includegraphics[width=0.9\columnwidth]{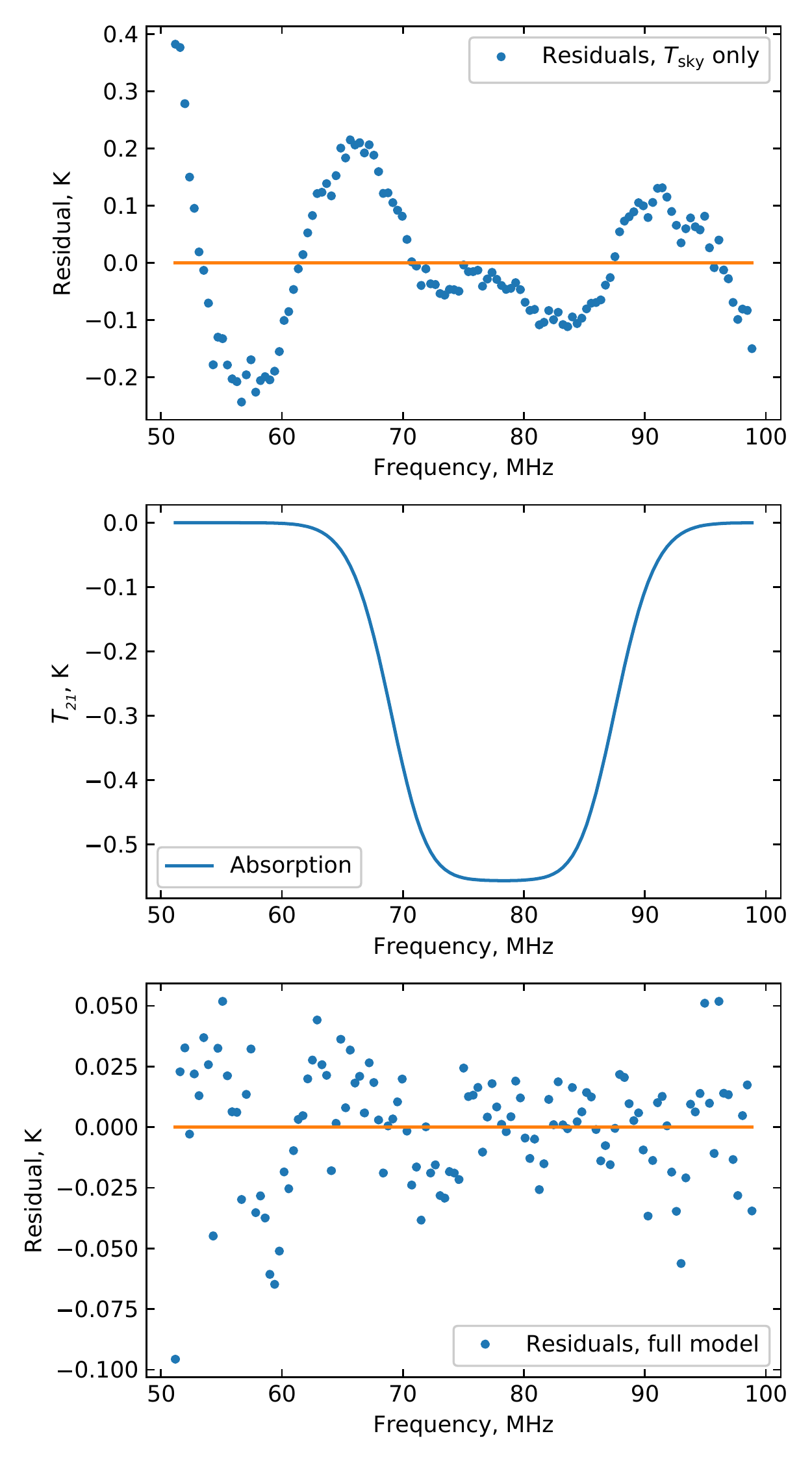}
    \caption{Best-fitting results, reproducing~\protect\citep{Bowman:2018yin}. \textit{Top:}~Residuals after fitting and removing only the foreground model, equation~(\protect\ref{eq:bowman-foreground}) (see their fig.~1b). 
    \textit{Middle:}~The best-fitting model of the 21-cm absorption feature, equation~(\protect\ref{eq:bowman-flgauss}) (see their fig.~1d).
    \textit{Bottom:}~Residuals after removing both the foreground and the 21-cm absorption model (see their fig.~1c).}
    \label{fig:bowman-weighted}
\end{figure}

Our best-fitting model closely reproduces the results reported by the EDGES collaboration. 
The best-fitting values for the 21-cm model parameters are $A=0.056$~K, $w=18.8$~MHz, and $\tau=5.8$. 
The resulting rms is 24.5~mK~\citep[close to 0.025~K reported by][]{Bowman:2018yin}. The summary of our results is plotted in Fig.~\ref{fig:bowman-weighted}. 
\section{Reproducing the original Bradley's results}\label{appendix:bradley}
\begin{figure}
    \centering
    \includegraphics[width=0.9\columnwidth]{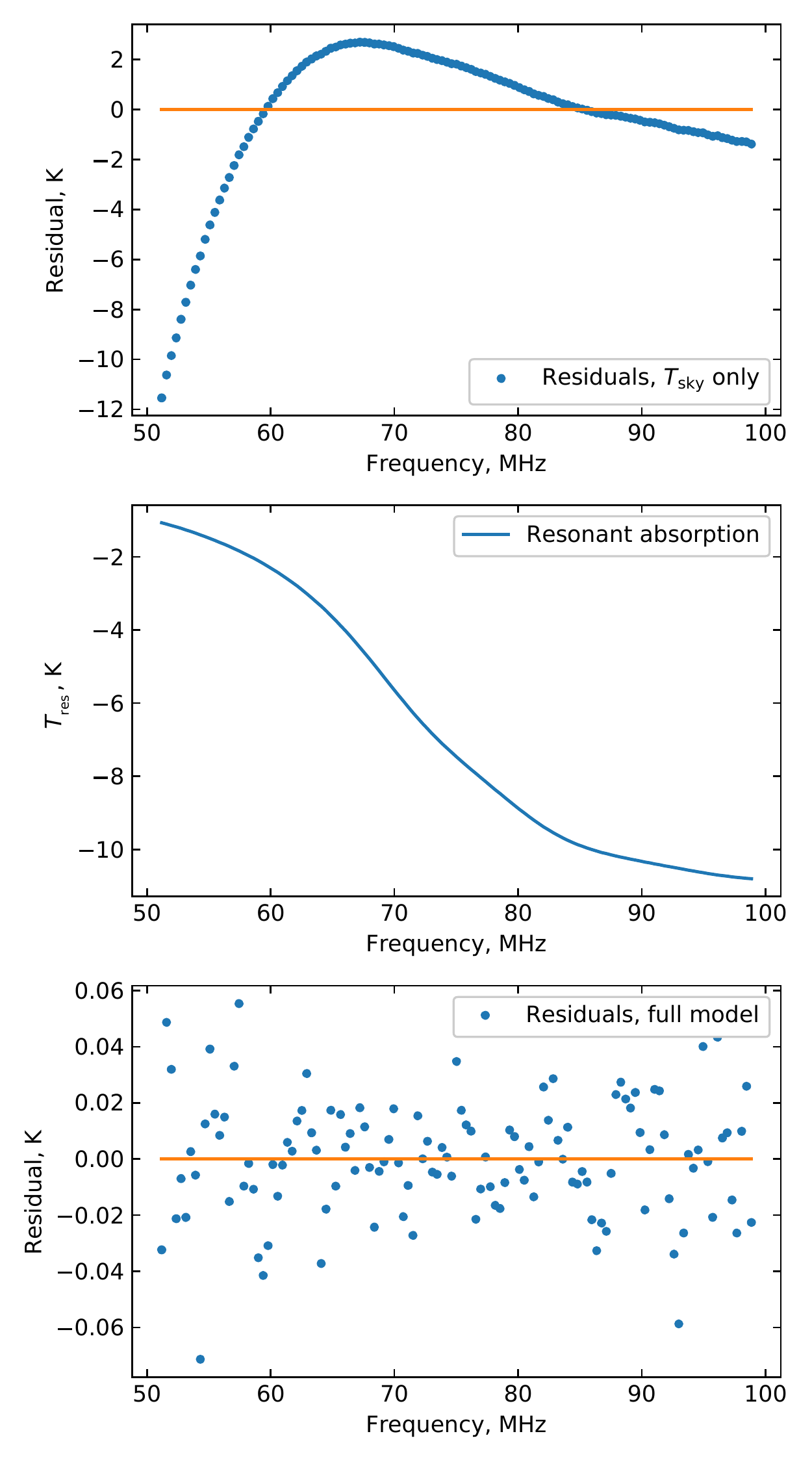}
    \caption{Best-fitting results, reproducing~\protect\citep{Bradley:2018eev}. \textit{Top:}~Residuals after fitting and removing only the foreground model (first two terms of equation~\protect\ref{eq:bowman-foreground}). 
    \textit{Middle:}~The best-fitting model of the instrumental resonant absorption feature (equation~\ref{eq:Tres}).
    \textit{Bottom:}~Residuals after removing both the foreground and the resonant absorption model.}
    \label{fig:bradley}
\end{figure}

We also reproduce the findings by \citep{Bradley:2018eev} so that the EDGES spectrum can be modelled with a sum of cosmic foreground continuum in the linearized form of equation~(\ref{eq:bowman-foreground}), taking into account only the first two terms, and three resonant features of the ground plane patch absorber (equation~\ref{eq:Tres}). The obtained best root mean square residual is 20.8~mK, 
the same as reported by~\citep{Bradley:2018eev}. The resonant absorption profile together with the model residuals are plotted in Fig.~\ref{fig:bradley}.

\section*{Data availability}
The EDGES low-band spectrum underlying this article is publicly available at: \url{http://loco.lab.asu.edu/edges/edges-data-release}.
The code that supports the findings of this study will be shared on reasonable request to the corresponding author.

%%%%%%%%%%%%%%%%%%%%%%%%%%%%%%%%%%%%%%%%%%%%%%%%%%



\bibliographystyle{mnras}
\bibliography{refs} % if your bibtex file is called example.bib

% Don't change these lines
\bsp	% typesetting comment
\label{lastpage}
\end{document}